PERSPECTIVES

# Empathic and agentic artificial intelligence in nursing: perspectives on a human-centered framework for cancer care navigation in the United States

**T. Girdwood[1]*, S. Kheirinejad[2], P. Kheirkhah[3], B. White[2], R. Davis[2], D. Schwartz[3] & A. Shaban-Nejad[2]**

[1]University of Tennessee Health Science Center, College of Nursing, Memphis; [2]University of Tennessee Health Science Center, Center for Biomedical Informatics, Department of Pediatrics, Memphis; [3]University of Tennessee Health Science Center, Department of Radiation Oncology, Memphis, USA

For patients experiencing cancer, nurse navigation can ease the burden of complex care by enhancing coordination of health services and patient outcomes. However, in under-resourced areas, trained nurse navigators may be limited or non-existent. In the United States, artificial intelligence (AI)-enabled digital health tools are increasingly available and may help address gaps in care coordination; however, most are not designed to specifically support nursing. This perspective piece discusses a human-centered AI framework that integrates empathic and agentic approaches grounded in the American Nurses Association's code of ethics to support nurses in the United States in cancer care navigation. The framework could augment, not replace, human empathy and agency while improving nurse workflow, patient—clinician relationships, and care coordination services in under-resourced areas.

Nurse navigation is a specialized service that addresses barriers to care, enhances care coordination across the cancer pathway, reduces healthcare costs, and improves patient satisfaction and health outcomes.[1] Trained nurse navigators play a crucial role in addressing whole-person patient needs like transportation to treatment sites and supporting physical, social, and emotional well-being.[1] However, navigators frequently lack institutional support, sustainable funding, and formal recognition of their specialized skillset.[2] In under-resourced areas in the United States (defined as settings lacking access to essential health resources like rural areas, communities experiencing significant health disparities, and underserved populations), patient access to specialty-trained nurse navigators may be limited or unavailable.

In this context, artificial intelligence (AI) offers a promising avenue to augment and support nurses who are not formally trained in navigation. AI-enabled digital health tools are maturing rapidly and can serve to supplement care access where there are fewer providers physically available, particularly in under-resourced regions.[3] Mobile health and virtual care tools can enhance nursing workflows, improve clinical decision making, and help ensure that patients in under-resourced areas experience fewer gaps in care and more coordinated health services.[4,5] These capabilities are especially important given that hospitalizations and readmissions associated with poor care coordination cost the United States healthcare system an estimated $27-78 billion annually.[6]

Despite their potential, most current AI-driven patient navigation systems prioritize automation and efficiency over patient priorities and emotional needs. This shortfall potentially undercuts individualized care needs, which are fundamental to the goals of nursing practice.[4] Currently, there is a lack of nursing-specific AI frameworks designed to align emerging technologies with ethical nursing principles such as empathy, advocacy, and patient-centered care.[4,5] This absence limits evaluation and improvement of AI tools and hinders insight into long-term impacts of nursing-specific AI frameworks on nurse clinical workflow and patient outcomes.[4,5,7]

## HUMAN-CENTERED AI CONCEPTUAL FRAMEWORK

We propose a human-centered AI conceptual framework that could support and empower nurses, particularly in under-resourced areas, by prioritizing individual patient needs, clinician—patient relationships, and care coordination services.[8] As illustrated in Figure 1, this framework integrates empathic and agentic AI capabilities while following principles in the American Nurses Association's (ANA) Code of Ethics.[10] The ANA Code of Ethics is the

**Correspondence to:* Dr Tyra Girdwood, The University of Tennessee Health Science Center, College of Nursing, 874 Union Ave, Memphis, TN 38163, USA. Tel: +1-901-448-9836

E-mail: tgirdwoo@uthsc.edu (T. Girdwood).

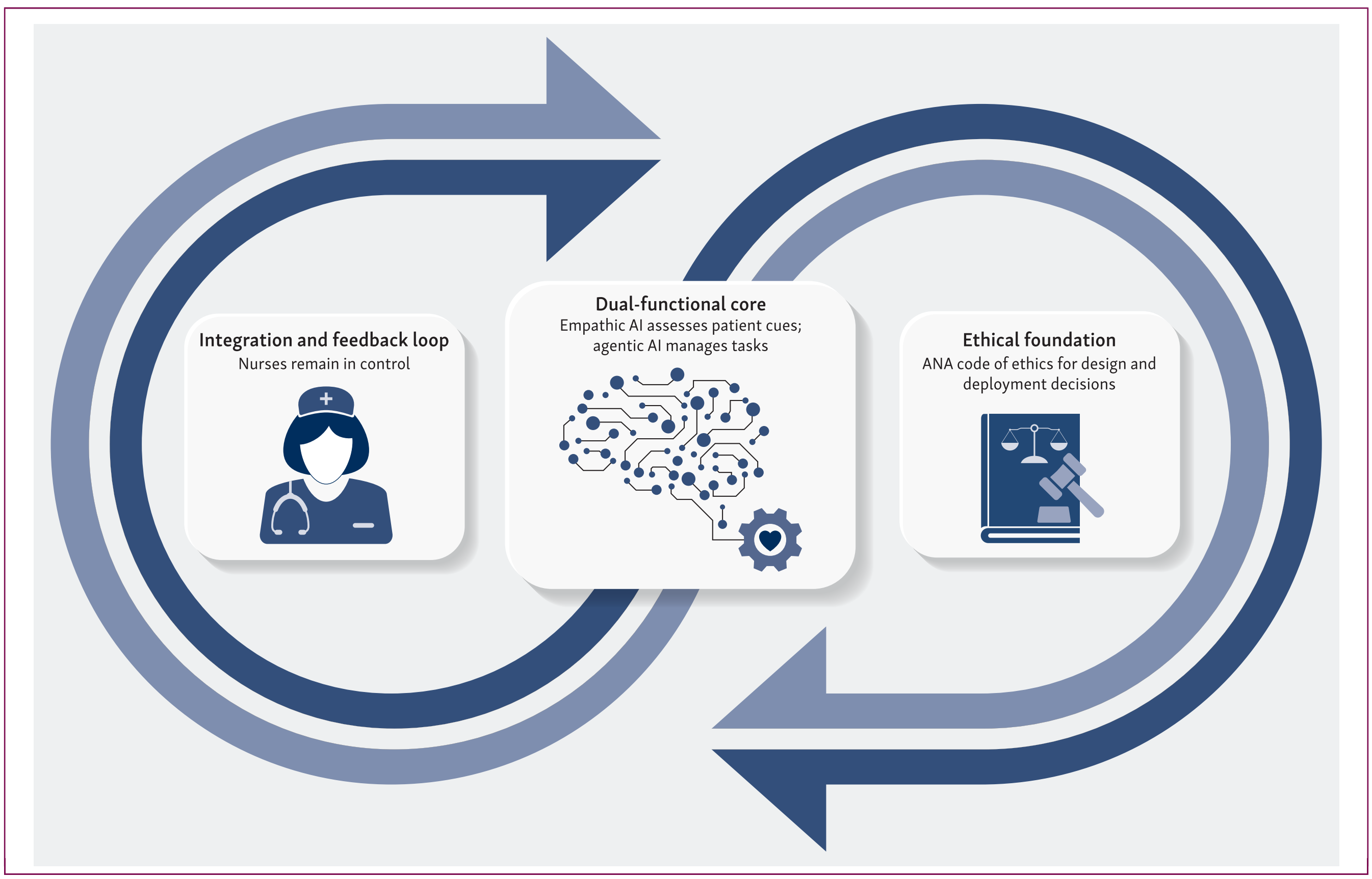

**Figure 1. Human-centered artificial intelligence (AI) in nursing framework (created by Use.AI[9]), showing ethical foundation, dual-functional core (empathic + agentic AI), and integration/feedback loop connecting nurses, patients, and technology.**

standard for ethical nursing practice and includes 10 provisions that encompass compassion, respect for human dignity, commitment to care, trust, advocacy, responsibility, accountability, and integrity.[10]

This framework consists of three mutually reinforcing layers, including: (i) strict compliance of all AI design, implementation, and governance decisions using the ANA Code of Ethics as a foundation; (ii) a dual-functional core where empathic AI modules detect and respond to patients' emotional and psychological cues while agentic AI modules proactively manage documentation, care coordination, information retrieval, and communication tasks; and (iii) intentional provider oversight designed to keep nurses firmly in control of all clinical decision-making that would directly impact patients. Together, these layers create a continuous, human-in-the-loop process where human empathy guides AI action and amplifies nursing capacity.

Empathic AI in healthcare refers to intelligent systems designed to recognize, interpret, and respond to patients' emotional, cognitive, and psychosocial states.[11] In nursing practice, empathic AI systems can alert nurses to a patient's emotional distress and help nurses facilitate communication and personalize care based on mood or stress indicators.[12] For example, AI-powered chatbots or virtual assistants can reach out to patients autonomously between clinical visits, detect early indicators of anxiety or depression, and escalate concerns so nurses can intervene proactively and efficiently.[13,14] Importantly, AI is not a substitute for human empathy, rather it serves to amplify nurses' empathic capacity by identifying subtle cues and automating supportive interactions without adding to their already demanding workloads.[15]

Agentic AI, by contrast, refers to systems capable of acting proactively and (semi) autonomously to execute tasks, generate recommendations, or coordinate care activities on behalf of nurses and multidisciplinary care teams.[16,17] In nursing practice, agentic AI could streamline administrative and operational duties such as documentation, care coordination, and scheduling.[18] Nurses currently dedicate substantial effort to charting and clerical tasks, limiting time for direct patient care while increasing risk for burnout.[19] By offloading repetitive low value tasks, agentic AI can reclaim time for direct patient interaction and education while improving the accuracy and consistency of nursing documentation and communication.

## BALANCING ENTHUSIASM WITH CAUTION

Cancer care navigation could uniquely benefit from novel integration of human-centered AI tools. Empathic and agentic AI capabilities promise to enhance coordination, communication, and continuity of care.[20] However, there remain crucial limitations and serious ethical risks that must be acknowledged and proactively managed.[20] Data privacy and security remain central concerns, as AI systems rely on large volumes of sensitive patient information like medical and social risk histories. Breaches, unauthorized access, or secondary data misuse could lead to significant patient harm, erosion of trust, and violation of regulatory standards like the Health Insurance Portability and Accountability Act. Robust safeguards, including data minimization, encryption, informed consent, audit trails, and role-based access controls are therefore essential components of any AI-enabled nursing workflow.[20] Equally concerning is the issue of algorithmic bias and inequitable system behavior. Without diverse representative and contextually relevant training data, empathic AI systems may misinterpret emotional cues across different cultural, linguistic or socioeconomic groups, while agentic AI may generate inaccurate, inequitable, or unsafe recommendations for marginalized populations.[5,21,22] To mitigate these risks, transparency, explainability, and accountability must be embedded throughout AI system design and deployment to monitor and challenge algorithmic outputs when necessary.[23]

Currently, there are no widely established clinical guidelines or legal precedents to clearly define accountability for AI-assisted decisions in nursing practice. This ambiguity raises important questions regarding professional responsibility, liability, and scope of practice. Moreover, the potential for dehumanization of care must be proactively addressed. Although empathic AI systems can simulate emotional understanding, they lack genuine empathy, moral reasoning, or relational awareness. Overreliance on automated interactions or AI-mediated communication could degrade patient—clinician relationships.[5] There also remain regulatory, legal, and ethical governance challenges for the safe application of AI.[20] Frameworks grounded in ethical principles, such as the ANA's Code of Ethics, offer a critical foundation for ongoing oversight, monitoring, transparency, and accountability to ensure that AI systems consistently support nursing values, professional integrity, and patient well-being.[24]

Integrating social data into AI care delivery support systems offers a powerful opportunity to personalize care, anticipate barriers, and identify patients at highest risk for treatment interruption or poor outcomes.[20] Continuous bias mitigation and rigorous fairness auditing must remain central priorities to prevent reinforcing disparities and inequities in care delivery.[5] Public health leaders, nursing organizations, policymakers, and community stakeholders should engage in participatory, community-informed work to establish 'equity-by-design' regulatory and governance frameworks that ensure transparency, fairness, and accountability are embedded at every stage of AI development and deployment. This is particularly critical in under-resourced areas where nursing and healthcare provider shortages persist.

## LOOKING TOWARD NURSING'S FUTURE

Before scaled adoption can occur, our proposed human-centered AI framework requires formal validation testing in clinical navigation settings to identify potential misalignments or errors, data privacy vulnerabilities, algorithmic biases, and workflow inefficiencies.[5] Our framework could be tested using existing AI-enabled navigation platforms (see Table 1) to assess and refine safety, usability, transparency, and ethical performance. Once rigorously tested and validated with nurses, providers, patients, and other stakeholders, this framework has the potential to advance nursing practice, leadership, and innovation across healthcare settings.

In real-world practice, our human-centered AI framework could help alleviate documentation burden, reduce cognitive overload, and enhance workflow efficiency across diverse cancer care settings and host communities.[5] Practical applications include automating responses to routine patient inquiries, assisting with triage decisions based on patient acuity, localizing and allocating community resources to address social drivers of health, and tailoring discharge instructions to patients' health literacy, language, and cultural context.

In nursing education, our framework can be integrated into virtual simulations, case-based learning, and training modules to teach undergraduate and graduate nursing students how to navigate AI-supported environments. Such integration would help students learn to critically evaluate AI outputs, address social drivers of health, and deliver coordinated empathic cancer care while maintaining ethical treatment principles and professional responsibility. Embedding this framework into nursing curricula would also promote digital literacy, ethical reasoning, and human-AI collaboration skills essential for the next generation of nurses.

Nurse scientists could apply our framework to systematically evaluate the long-term impact of AI integration on provider well-being, patient outcomes, care quality, and organizational performance across different settings. This evidence base would guide best practices and help refine human-centered AI systems for equitable and safe implementation. Before widespread adoption, professional organizations such as the ANA should establish policies and standards that govern AI integration within nursing practice. These guidelines should address ethical consistency, clinical accountability, and liability in AI-assisted care to ensure nurses retain professional autonomy and responsibility for patient outcomes. Establishing such guidance is essential to protect patient safety and prevent unintended harm to patients, their families, and healthcare providers as AI becomes increasingly embedded in clinical care. Future work is needed to examine how our human-centered AI framework may translate to non-United States healthcare systems with different workforce models, regulatory environments, and care delivery structures.

**Table 1. Examples of oncology AI patient navigation platforms or tools**

| Platform or tool | How AI supports patient navigation | Reference |
|---|---|---|
| Care.AI | Virtual nursing and smart room platform that automates routine monitoring and documentation, and supports virtual nursing models to manage high-acuity oncology patients | Care.AI. The world's most advanced AI-assisted virtual nursing platform. n.d. Available at https://www.care.ai/virtual-nursing.html. Retrieved 1 February 2026. |
| Fidari | Platform streamlines standardized oncology care pathways, automates task assignment (agentic AI), stratifies patients based on risk, and tasks are assigned to the right health care provider | Fidari. Ensure your patients navigate care with confidence and support. n.d. Available at https://fidari.care/providers. Retrieved 1 February 2026. |
| MyEleanor (Montefiore Einstein Comprehensive Cancer Center) | AI-based virtual patient navigator that finds and reengages oncology patients overdue for screenings | Pratt-Chapman M. Navigation refresh: AI and oncology patient navigation. *J Oncol Navig Surviv.* 2025 September;16 (9). Available at https://www.jons-online.com/issues/2025/september-2025-vol-16-no-9/navigation-refresh-ai-and-oncology-patient-navigation. |
| iNAV (Northwell Health) | AI tool that analyzes MRI and CT scans to detect cancers earlier; when flagged, care navigators expedite biopsies and treatment planning | Burroughs A. AI Support in the patient care journey. Health Tech Magazine; 2025 May. Available at https://healthtechmagazine.net/article/2025/05/ai-support-patient-care-journey. |
| Medical University of South Carolina Health | AI platform that reviews medical records, detects patients overdue for mammograms, and sends these patients a self-scheduling link for their screening | Burroughs A. AI Support in the patient care journey. Health Tech Magazine; 2025 May. Available at https://healthtechmagazine.net/article/2025/05/ai-support-patient-care-journey. |
| AI-enabled tools (Generally available) | (i) AI-enhanced patient portals<br>(ii) Chatbots for symptom triage and coordinating patient appointments<br>(iii) mHealth apps that support patient—clinician communication and follow-up<br>(iv) AI-assisted documentation and referral tools | Enhancing patient navigation with technology to improve equity in cancer care: a report to the President of the United States from the President's Cancer Panel. Bethesda (MD); 2024. Available at https://prescancerpanel.cancer.gov/reports-meetings/enhancing-patient-navigation-2024. |

AI, artificial intelligence; CT, computed tomography; MRI, magnetic resonance imaging.

## DECLARATION OF GENERATIVE AI AND AI-ASSISTED TECHNOLOGIES IN THE WRITING PROCESS

During the preparation of this work the authors used 'Use.AI' (with detailed instructions written by the authors) to create a visual of our framework (Figure 1). After employing this tool, the authors reviewed and edited the image as needed and take full responsibility for this content.

## ACKNOWLEDGEMENTS

The authors extend their gratitude to colleagues whose insights supported the development of this perspective piece.

## FUNDING

None declared.

## DISCLOSURE

The authors have declared no conflicts of interest.